\documentclass[conference]{svproc}
\usepackage{cite}
\usepackage{graphicx}
\usepackage{amsmath,amssymb,amsfonts,booktabs,tabularx}
\usepackage{algorithmic}
\usepackage{array}
\usepackage{hyperref}
\usepackage[pagewise]{lineno}
\usepackage{tikz}
\usetikzlibrary{arrows.meta, positioning}

\begin{document}

\title{Review of Tools for Zero-Code LLM Based Application Development}



\author{Priyaranjan Pattnayak\inst{1} \and Hussain Bohra\inst{2}}
\authorrunning{Pattnayak et al.} 
%
%
\institute{University of Washington, Seattle, USA,\\
\email{ppattnay@uw.edu},
\and
SVKM's Narsee Monjee Institute of Management Studies, Mumbai, India,\\
\email{hussainbohra@gmail.com}
}
\maketitle

\begin{abstract}
Large Language Models (LLMs) are transforming software creation by enabling zero-code development platforms. This survey reviews recent platforms that let users build applications without writing code, by leveraging LLMs as the “brains” of the development process. We adopt a broad survey methodology, categorizing platforms based on key dimensions such as interface style, backend integration, output type, and extensibility. We cover both dedicated LLM-based app builders (e.g., OpenAI’s custom GPTs, Bolt.new, Dust.tt, Flowise, Cognosys) and general no-code platforms (e.g., Bubble, Glide) that integrate LLM capabilities. We present a taxonomy categorizing these platforms by their interface (conversational, visual, etc.), supported LLM backends, output type (chatbot, full application, workflow), and degree of extensibility. Core features such as autonomous agents, memory management, workflow orchestration, and API integrations are examined. We provide a detailed comparison, highlighting each platform’s strengths and limitations. Trade-offs (customizability, scalability, vendor lock-in) are discussed in comparison with traditional and low-code development approaches. Finally, we outline future directions—including multimodal interfaces, on-device LLMs, and improved orchestration for democratizing app creation with AI. Our findings indicate that while zero-code LLM platforms greatly lower the barrier to creating AI-powered applications, they still face challenges in flexibility and reliability. Overall, the landscape is rapidly evolving, offering exciting opportunities to empower non-programmers to create sophisticated software.
\end{abstract}


\section{Introduction}
The rise of no-code and low-code development platforms has significantly lowered the barrier for non-programmers to create software \cite{noCodeSurvey2021}. By providing visual interfaces, templates, and pre-built components, these platforms have democratized app development, enabling “citizen developers” to build and deploy applications without traditional coding \cite{noCodeForAll2023}. However, until recently, the complexity of custom logic still posed challenges—anything beyond simple workflows often required some coding or technical expertise. 

Large Language Models (LLMs) have emerged as a game-changer in this context by allowing natural language to be used as a programming interface. Modern LLMs like GPT-4 can interpret high-level instructions and even generate working code \cite{openai2023gpts, liu2023gptsurvey, touvron2023llama}. This capability opens the door for zero-code platforms where a user’s plain-English description is enough to create functional software prototypes\cite{ouyang2022training}. In essence, LLMs enable natural language-driven development, turning vague ideas into functional prototypes faster than ever \cite{ziegler2023instruction}. The prospect is that anyone can describe an application or process, and an AI system will handle the implementation details – further democratizing software creation beyond what earlier no-code tools achieved. 

Several new platforms now leverage LLMs to enable application building with minimal or no coding. Some are dedicated LLM-based app creators, built from the ground up around LLM capabilities. For example, OpenAI’s custom GPTs allow users to create tailored chatbots by providing instructions and knowledge, “no coding is required” \cite{openai2023gpts}. Other platforms like Bolt.new let users converse with an AI agent that writes and deploys code for full-stack applications \cite{bolt2023gettingstarted}. Similarly, platforms such as Dust.tt \cite{dustDocs2023} and Cognosys \cite{cognosys2023} focus on creating AI agents and workflows through natural language and a friendly UI. In parallel, traditional no-code platforms are adding LLM integrations. Tools like Bubble \cite{bubble2023plugin} and Glide \cite{glide2023ai} now make it easy to plug AI services (e.g., OpenAI’s GPT-4) into apps built with their visual editors, enabling AI features such as text generation, summarization, and chat within no-code apps.

This paper provides a comprehensive survey of these zero-code LLM platforms. We first establish a taxonomy to categorize the diverse approaches. We then examine core features and compare representative platforms side-by-side, using tables to summarize capabilities. We discuss key trade-offs—such as flexibility vs. simplicity, and the risks of vendor lock-in or scalability limits—when adopting these platforms. A comparison with traditional software development and low-code approaches is included to highlight what challenges remain. Finally, we identify future directions and emerging trends (multimodal interfaces\cite{pattnayak2024survey}, on-device models, better AI orchestration) that could further enhance no-code AI app creation. While comprehensive, this survey does not follow a traditional review protocol but focuses on representative, influential platforms available publicly. The goal is to inform practitioners, researchers, and prospective “no-code developers” about the current landscape, its benefits, and its limitations.

\section{Taxonomy of Zero-Code LLM Platforms}
The ecosystem of zero-code LLM platforms can be categorized along several dimensions. Figure~\ref{fig:taxonomy} presents a taxonomy summarizing two primary classes\textemdash platforms dedicated to LLM-based development vs. general-purpose no-code platforms \textemdash and further subcategories based on interface and design paradigm. We discuss four key taxonomic dimensions: interface type, LLM backend support, output type, and customization level as mentioned in Table \ref{tab:taxonomy-pivot}~\cite{dustDocs2023, flowiseDocs, cognosys2023, bubble2023plugin, glide2023ai, openai2023gpts, bolt2023gettingstarted, anthropic2024claude, lewis2020rag}.
\begin{table}[t]
\centering
\caption{Taxonomy Summary: Dedicated vs. General Zero-Code Platforms}
\renewcommand{\arraystretch}{1.2}
\begin{tabular}{p{3cm} p{4.5cm} p{4cm}}
\toprule
\textbf{Dimension} & \textbf{Dedicated LLM Platforms} & \textbf{General No-Code Platforms} \\
\midrule
\textbf{Interface Type} & Chat, Visual Flows & GUI Builder, Templates \\
\textbf{Output Type} & Agent, Workflow & Full App, Embedded AI \\
\textbf{LLM Backend} & Model-agnostic, OpenAI, Anthropic & Mostly OpenAI, Auto-select \\
\textbf{Customizability} & Low-code hooks, SDKs & Pure no-code, Plugin mode \\
\textbf{Examples} & GPTs, Flowise, Dust.tt & Bubble, Glide \\
\bottomrule
\end{tabular}
\label{tab:taxonomy-pivot}
\end{table}

\begin{itemize}
    \item \textbf{Platform Category} -- Some platforms are purpose-built for LLM applications, whereas others are general no-code/low-code platforms extending their features with LLM integration. Dedicated LLM platforms (e.g., Dust.tt, Flowise, Cognosys) are often geared towards creating AI-driven agents or workflows. In contrast, general platforms (e.g., Bubble, Glide) focus on broader application development (web/mobile apps, internal tools) and incorporate LLMs as an add-on for AI features. This categorization often correlates with the target use-cases: LLM-specific platforms excel at conversational agents, text analysis, and task automation, while general no-code tools cover use cases like form-based apps or database-driven apps, now augmented with AI capabilities.

    \item \textbf{Interface Type} -- Zero-code LLM platforms offer different user interfaces for building applications:
    \begin{itemize}
        \item \textit{Conversational (Chat-based):} The user “chats” with an AI agent to build the app or agent. The platform interprets natural language instructions. For example, OpenAI’s custom GPTs are created by a dialog where the user provides instructions and examples to the GPT agent~\cite{openai2023gpts}. Bolt.new similarly provides a chat-based development environment, where the user tells the AI what to build or change, and the AI writes code accordingly~\cite{bolt2023gettingstarted}. This interface is highly accessible: the design process feels like instructing a smart assistant.
        \item \textit{Visual Programming (Flow/Graph):} The user assembles the application logic via a drag-and-drop editor or diagram. Flowise is an example offering a visual flow builder: users connect nodes representing LLMs, tools, or data sources to define an AI workflow (similar to Node-RED or BPMN for AI)~\cite{flowiseDocs}. This is suited for users who prefer a schematic overview of the AI reasoning process (e.g., adding a node for memory or an API call).
        \item \textit{Form \& Template Configuration:} The user assembles the application logic via a drag-and-drop editor or diagram. Flowise is an example offering a visual flow builder: users connect nodes representing LLMs, tools, or data sources to define an AI workflow (similar to Node-RED or BPMN for AI). This is suited for users who prefer a schematic overview of the AI reasoning process (e.g., adding a node for memory or an API call).
        \item \textit{Traditional No-Code Builder:} General platforms like Bubble and Glide present a GUI builder (with widgets, property panels, workflow rules). Here, LLM integration might be configured via actions or plugin settings. For example, in Bubble, a user might drag a text element and set its content to “AI response”, wiring it to an OpenAI API call via a visual workflow~\cite{bubble2023plugin}. Glide’s approach is to allow AI features in its Data Editor or Action Editor, where the user specifies what kind of AI transformation to apply to data (e.g., “extract date from text using AI”)~\cite{glide2023ai}.
    \end{itemize}

    \item \textbf{LLM Backend} -- The choice of underlying LLM(s) varies:
    \begin{itemize}
        \item Some platforms are tied to a specific model or provider. OpenAI’s GPT-based platform naturally uses OpenAI’s models (GPT-3.5, GPT-4)~\cite{openai2023gpts}. Cognosys and Bolt.new have primarily used OpenAI or Anthropic models – for example, Bolt leverages Anthropic’s Claude for code generation~\cite{bolt2023gettingstarted, anthropic2024claude}. If a platform is closed-source SaaS, users may not even know exactly which model is powering it (though the platform may offer options like “GPT-4 vs GPT-3.5”).
        \item Other platforms are model-agnostic or support multiple LLMs. Flowise explicitly supports integration with various models and AI services; it lists compatibility with LangChain, LlamaIndex, and 100+ integrations including different LLM providers~\cite{flowiseDocs}. Dust.tt also emphasizes a model-agnostic approach, aiming to let users leverage “the best AI models for each task” and avoid lock-in. Such platforms often allow switching between providers (OpenAI, Anthropic, etc.) or even using open-source models, which can be beneficial for cost and flexibility.
        \item An emerging aspect is on-device or private LLM support. Currently, most no-code platforms rely on cloud APIs for LLM access due to the resource intensiveness of models. However, as smaller efficient models (e.g., Llama 2 variants) become available, future platforms might let users run LLMs locally or in a private cloud for data privacy~\cite{lewis2020rag}. We discuss this trend in Future Directions (Section on Future Trends).
    \end{itemize}
\begin{figure*}[t!]
  \centering
  \includegraphics[width=\textwidth]{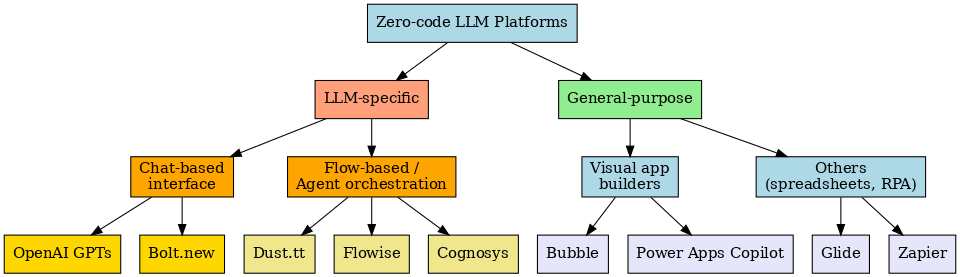}
  \caption{Taxonomy of zero-code platforms for LLM-powered app creation.}
  \label{fig:taxonomy}
\end{figure*}
    \item \textbf{Output Type} -- Zero-code LLM platforms differ in what they ultimately produce or deploy:
    \begin{itemize}
        \item \textit{Chatbot or AI Assistant:} Many LLM-specific platforms create an AI agent that interacts in natural language. OpenAI GPTs are essentially custom chatbots accessible through the ChatGPT interface~\cite{openai2023gpts}. Dust.tt focuses on deploying agents that can be chatted with or invoked via API (e.g., in Slack or on a website)~\cite{dustDocs2023}. Cognosys builds autonomous agents that execute tasks and present results, often through a chat-like interface that shows the agent’s thought process~\cite{cognosys2023}. In these cases, the end-product is an AI assistant, not a traditional UI-based app.
        \item \textit{Full Applications (Web/Mobile):} Some platforms generate complete applications with user interfaces. Bolt.new, for example, can create and deploy a full-stack web or mobile app – generating the code (HTML/CSS/JS, backend) and hosting it via integrations like Netlify~\cite{bolt2023gettingstarted}. The user ends up with a working web app they can use or further refine. General no-code platforms like Bubble output web applications; when enhanced with LLM logic, the output remains a web app but now with AI-driven features (like a chatbot widget, or AI-generated content in the UI)~\cite{bubble2023plugin}.
        \item \textit{Workflow/Automation:} Some outputs are behind-the-scenes workflows or API endpoints rather than user-facing apps. For instance, Dust.tt’s “Dust Apps” enable orchestrating backend workflows (LLM calls, data queries, API calls) that could be invoked via an API~\cite{dustDocs2023}. Similarly, one could use a platform like Flowise or Zapier with LLM integration to build an automation pipeline (no UI except maybe a simple interface to trigger or monitor it). These outputs integrate into other systems or run background processes.
        \item \textit{Hybrid:} A few platforms blur the lines – e.g., Glide produces an app (often mobile or web) but one of the components within it could be an AI chatbot~\cite{glide2023ai,pattnayak2025hybrid}. So the output is an application with an AI component. Likewise, Power Apps with AI can create a business app that also has a Copilot chat for queries.
    \end{itemize}

    \item \textbf{Customization and Extensibility} -- Another critical dimension is how much one can extend or modify beyond the provided no-code options:
    \begin{itemize}
        \item \textit{Pure No-Code:} Platforms like OpenAI’s GPT builder or Cognosys aim for zero coding; users cannot (and need not) write any code~\cite{openai2023gpts, cognosys2023}. Customization is through natural language instructions or configuration only. This makes them very accessible, but potentially limiting if the desired behavior isn’t achieved via provided options.
        \item \textit{Low-Code Hooks:} Some “no-code” tools allow optional code for advanced users. Flowise, while primarily drag-and-drop, is developer-friendly—one can create custom tool plugins or write scripts to extend its functionality~\cite{flowiseDocs}. Dust.tt provides a no-code interface but also a developer API and the ability to embed agents in custom apps~\cite{dustDocs2023}.
        \item \textit{Exportability:} A form of extensibility is whether the platform lets you export the generated code or model prompts for use outside the platform. Bolt.new, for example, shows you the live code it’s writing and runs it in a container; an advanced user could take that code and continue development off-platform~\cite{bolt2023gettingstarted}. In contrast, proprietary platforms might keep the logic hidden—e.g., you can design an agent in Cognosys, but you cannot directly extract all the underlying code/logic.
        \item \textit{Integration:} Extensibility also means how well the platform integrates with external services. Platforms differ on whether they allow adding new data sources or connecting third-party APIs without code. Dust.tt advertises “no ceiling on data connections” so that users can connect any knowledge base or tool to their agent~\cite{dustDocs2023}. General platforms like Bubble have plugin systems (Bubble’s API connector and plugin marketplace) enabling integration with any external AI service or database~\cite{bubble2023plugin}. Glide similarly allows integrating OpenAI or other AI via API if something isn’t available natively~\cite{glide2023ai}.
    \end{itemize}
\end{itemize}

This taxonomy helps clarify that “no-code LLM platform” is not a monolith—there is a spectrum from highly specialized chat-based agent builders to general app builders embedding AI. Each category comes with a different user experience and caters to different user needs. In the next sections, we examine specific platform examples through the lens of these categories and discuss their core capabilities.

\section{Core Features and Capabilities}
Beyond the broad categories above, zero-code LLM platforms can be compared by the concrete features they offer. These features determine what kinds of applications or agents can be built and how intelligent or complex they can be. We highlight six key capabilities: support for autonomous agents (tool use and multi-step reasoning), memory and knowledge integration, workflow logic and control, integration with external services via APIs, and other notable features (like multimodal inputs or training customization)~\cite{dustDocs2023, flowiseDocs, cognosys2023, openai2023gpts, bolt2023gettingstarted, glide2023ai, bubble2023plugin, microsoft2023copilot}. A detailed comparision is shown in Table \ref{tab:comp_p1} and Table \ref{tab:comp_p2}.

\subsection{Agent Support (Autonomy and Tool Use)}
Many platforms enable the creation of autonomous agents that break down tasks and invoke APIs. Cognosys acts like a web-based AutoGPT, executing task loops iteratively~\cite{cognosys2023}. Dust.tt supports chaining model calls and tool execution~\cite{dustDocs2023}. Flowise includes agent modules with tool integration~\cite{flowiseDocs}. In contrast, OpenAI GPTs provide limited autonomy~\cite{openai2023gpts}, and general tools like Bubble or Glide require users to manually sequence agent-like behavior~\cite{bubble2023plugin, glide2023ai}.

\begin{table*}[h!]
\centering
\caption{Comparison of Core Features of Zero-Code LLM Platforms (Part 1)}
\renewcommand{\arraystretch}{1.2}
\begin{tabular}{l p{2.5cm} p{3.8cm} p{4cm}}
\toprule
\textbf{Platform} & \textbf{Interface Style} & \textbf{Agent/Tool Use} & \textbf{Memory} \\
\midrule
OpenAI & Chat & Limited tools & Session + setup knowledge \\
Bolt.new & Chat + code & Full API + tool use & Session + conversation memory \\
Dust.tt & Templates + flow & Full agent toolkit & RAG/data integration \\
Flowise & Visual graph & Custom agents/tools & Vector DBs via LangChain \\
Cognosys & Chat + agent UI & Autonomous agent loops & Partial (search, Q\&A) \\
Bubble & Visual builder & No agents (uses API) & DB/plugins for memory \\
Glide & Visual builder & No agents (embed chat) & AI field-level memory \\
\bottomrule
\end{tabular}
\label{tab:comp_p1}
\end{table*}

\begin{table*}[t]
\centering
\caption{Comparison of Core Features of Zero-Code LLM Platforms (Part 2)}
\renewcommand{\arraystretch}{1.2}
\begin{tabular}{l p{4cm} p{2.5cm} p{3.2cm}}
\toprule
\textbf{Platform} & \textbf{Workflow Logic} & \textbf{LLM Backend} & \textbf{Extensibility} \\
\midrule
OpenAI GPTs & None (chat-driven) & GPT-only & Closed (no-code only) \\
Bolt.new & Guided via chat & Claude & Exportable code \\
Dust.tt & Chain + conditional logic & Multiple & API + low-code \\
Flowise & Branches, loops via nodes & Any (open-source) & Fully extensible \\
Cognosys & Agent-decided logic & GPT-based (likely) & Closed SaaS \\
Bubble & Classical workflow engine & Any via API & Plugins, JS \\
Glide & Event-driven logic & OpenAI & Formulas, limited code \\
\bottomrule
\end{tabular}
\label{tab:comp_p2}
\end{table*}

\subsection{Memory and Knowledge Integration}
Effective LLM applications often need memory of past interactions or access to domain knowledge. Platforms address this in different ways:
\textbf{Short-Term Memory:} In chat-based builders, the conversation history serves as memory. For example, OpenAI GPTs include the prior dialogue automatically, allowing the custom GPT to maintain context within a session. Bolt.new, while building an app via chat, effectively has the conversation log to remember user instructions and previous code it wrote. Ensuring the model doesn’t forget earlier instructions is crucia`l; some platforms may behind-the-scenes use prompt engineering or store intermediate state to remind the model of the project specifications after each response.

\textbf{Long-Term Knowledge:} Some platforms allow uploading documents~\cite{meghwani2025hardnegativeminingdomainspecific} or connecting data sources that the AI can reference. This is common in enterprise-focused platforms like Dust.tt, which emphasizes connecting internal knowledge bases (Slack, Google Drive, Notion, etc.) so that agents can answer with company-specific information. Dust.tt uses Retrieval Augmented Generation (RAG) techniques – indexing data so that relevant context is fetched and fed into the prompt when the agent answers. Flowise, being based on LangChain, also makes it easy to add vector databases or data loaders for knowledge retrieval. Even general no-code tools have started offering this: for instance, Bubble has plugins for vector search or one could integrate a service like Pinecone to store and retrieve embeddings~\cite{bubble2023plugin}. Glide’s Glide AI supports “Instructions” and can incorporate existing data from the app’s tables as context, though for more advanced memory a developer might use external services~\cite{glide2023ai}.

\textbf{Custom Instructions:} OpenAI’s GPT builder allows the creator to give extra knowledge or instructions to the custom GPT, which can be seen as a form of one-time prompt memory every time that GPT is used. Some platforms also support fine-tuning or at least saving conversation state. Generally, “memory” is either implicit (chat history) or explicit (vector stores, databases). The more a platform targets complex tasks, the more likely it has an explicit memory/knowledge integration feature.

\subsection{Workflow and Control Logic}
Traditional no-code apps often need conditional logic, loops, or event-driven triggers. Similarly, LLM-based apps sometimes require orchestrating multiple steps or handling branching depending on AI outputs. Platforms like Dust.tt treat a whole sequence of model calls and tool calls as a single app specification, where you can orchestrate complex branching workflows. For example, a Dust app could: take user input, call an LLM to analyze it, if the LLM indicates a need for data lookup then call a database, then feed results to another LLM call, etc. This is essentially workflow orchestration with AI in the loop. Flowise explicitly supports branching and looping by virtue of its node-graph interface ~\cite{flowiseDocs}—one can add decision nodes based on LLM outputs or error-handling nodes.

In contrast, chat-centric platforms might not expose explicit branching logic to the user; they rely on the LLM’s reasoning within a single prompt. For instance, you might instruct a GPT “If the user asks about X, first do Y then Z” in natural language rather than using an if-else block. This can be less predictable. General no-code platforms allow classical workflows (e.g., Bubble’s workflow engine) and one can incorporate AI decisions by parsing the AI’s response. However, this often requires creative use of the tool (like expecting the LLM to return a specific keyword that you then use in a conditional step). We observe that platforms like LangFlow/Flowise or Dust—influenced by AI developer frameworks—provide more robust control flow constructs for AI-driven logic, whereas pure no-code solutions lean on the LLM’s internal decision-making or simpler sequential triggers~\cite{bubble2023plugin}.

\subsection{API Integration and Tool Connectivity}
A powerful LLM agent often needs to use external tools (e.g. call an API, query a database, send an email). Many no-code LLM platforms incorporate this via “tools” or integration modules:
\begin{itemize}
  \item \textbf{Dust.tt:} Call APIs, execute code, trigger agents programmatically~\cite{dustDocs2023}.
  \item \textbf{Flowise:} Supports custom tools via LangChain (e.g., weather APIs)~\cite{flowiseDocs}.
  \item \textbf{Cognosys:} Auto-selects tools like search, file Q\&A~\cite{cognosys2023}.
  \item \textbf{General no-code platforms:} Bubble, Glide, and Zapier connect to APIs using plugins or visual editors~\cite{bubble2023plugin, glide2023ai}.
\end{itemize}
The bottom line is that the breadth of integrations a platform supports greatly determines its potential use cases. Platforms positioned for business use (Dust, Cognosys) advertise connectivity to internal tools and data; open-source projects (Flowise) leverage community-contributed integrations; app builders (Bubble, Glide) leverage their existing plugin ecosystems.

\subsection{Multimodal and AI-Assisted Features}
Some platforms go beyond text. For example, Dust.tt and Cognosys have experimental support for image analysis and other modalities~\cite{dustDocs2023, cognosys2023}. Glide AI allows extracting information from images (like scanning a receipt to extract expense data) and even audio transcription via integrated AI services. These multimodal capabilities \cite{patel2024llm,pattnayak9339review,agarwal2024mvtamperbench}are increasingly important for real-world tasks (documents often contain images or need OCR, etc.)\cite{pattnayak2025clinicalqa20multitask,pattnayak2025tokenizationmattersimprovingzeroshot,pattnayak2025tokenization}. Another feature to note is AI-assisted development: Bubble recently introduced an AI assistant that can help build the app (Bubble’s own “AI Builder” that generates pages from a description) — effectively using an LLM to assist within the no-code editor\cite{bubble2023plugin}. Microsoft’s Power Apps Copilot \cite{microsoft2023copilot} similarly “generates the app for you based on your description”, creating screens and a data model from a prompt. While not a feature of the built app itself, this use of LLMs inside the development process is a noteworthy capability that blurs the line between no-code and automatic code generation. We discuss such assistance in comparison to traditional development later.

\begin{itemize}
  \item \textbf{Multimodal:} Dust.tt, Cognosys, and Glide handle images/audio/text.
  \item \textbf{AI Assistants:} Bubble's AI Builder and Microsoft Power Apps Copilot generate UI/screens from prompts.
\end{itemize}
These enhance app intelligence and developer productivity.

\section{Detailed Platform Comparisons}
To illustrate the landscape, we provide a comparison, as shown in Table \ref{tab:feature-comparison}, of several notable platforms, expanding on their workflows and distinguishing characteristics~\cite{openai2023gpts, bolt2023gettingstarted, dustDocs2023, flowiseDocs, cognosys2023, bubble2023plugin, glide2023ai}.

\subsection{OpenAI Custom GPTs}
OpenAI's Custom GPTs, launched in late 2023, enable users to configure chatbots without code~\cite{openai2023gpts}. Users can provide instructions, upload optional knowledge bases, and enable tools like web browsing or code execution. These bots live within the ChatGPT ecosystem and are accessible via a dedicated UI. They are easy to configure and ideal for targeted tasks, such as a Travel Planner GPT, but are limited in logic complexity and autonomy.

\subsection{Bolt.new}
Bolt combines a chat interface with real-time code generation using Claude~\cite{bolt2023gettingstarted}. It enables natural language prompts like \textit{"Build a to-do app"}, which are converted into React/Node projects. The platform supports editing, previewing, and deployment via Netlify. It suits users with some coding familiarity who benefit from both AI assistance and full-code access. Limitations arise when debugging or prompt clarity is needed.

\begin{table}[t]
\centering
\caption{Comparison of Zero-Code Platforms by Feature Support}
\renewcommand{\arraystretch}{1.2}
\setlength{\tabcolsep}{4pt}
\begin{tabular}{p{3cm} c c c c c}
\toprule
\textbf{Feature} & \textbf{GPTs} & \textbf{Cognosys} & \textbf{Flowise} & \textbf{Zapier AI} & \textbf{Dust.tt} \\
\midrule
LLM Flexibility & Medium & High & High & Low & Medium \\
Workflow Logic  & Medium & High & High & High & Medium \\
API Integration & Medium & Low & High & High & Medium \\
UI Generation   & Low & Medium & Low & None & Medium \\
Agent Support   & Medium & High & Medium & Low & Medium \\
Open Source     & No & No & Yes & No & Partial \\
\bottomrule
\end{tabular}
\label{tab:feature-comparison}
\end{table}

\subsection{Dust.tt}
Dust offers a platform to build LLM-powered agents and workflows~\cite{dustDocs2023}. Users can construct "Dust apps" composed of prompt chains, retrieval steps, and API/tool calls. It emphasizes data connectivity, enterprise use, and agent monitoring. While accessible to non-programmers via templates, it also provides SDKs for developers. Its design promotes flexibility and vendor-independence.

\subsection{Flowise (Open-Source Builder)}
Flowise is a visual editor built on LangChain~\cite{flowiseDocs}. It enables building flows using nodes (e.g., document loader, vector store, LLM Q\&A). It supports deployment as a chat widget and offers open-source self-hosting for privacy and cost-efficiency. Flowise is favored by developers seeking quick prototyping with visual debugging, though its interface may overwhelm non-technical users.

\subsection{Cognosys}
Cognosys focuses on AutoGPT-style agents for non-technical users~\cite{cognosys2023}. The platform abstracts complexity, letting users define objectives and execute task loops asynchronously. It supports integrations like Google Drive and Gmail. While ideal as a personal AI assistant, it is closed-source and constrained to pre-defined agent modes. Customizability is low but ease-of-use is high.

\subsection{Bubble with AI Integration}
Bubble is a general-purpose no-code builder enhanced with AI capabilities~\cite{bubble2023plugin}. Developers can connect OpenAI APIs to generate or process text within workflows. Bubble’s plugin ecosystem supports model selection and prompt configuration. Additionally, its new "AI Assistants" feature accelerates app construction via prompts. Unlike agent-specific tools, Bubble builds complete multi-page apps with embedded AI elements.

\subsection{Glide}
Glide turns spreadsheets into apps and incorporates AI via Glide AI~\cite{glide2023ai}. It focuses on transforming text/images/audio into structured data fields. Users can extract information from receipts, emails, or voice notes. It’s optimized for internal tools and excels in structured automation tasks, though it lacks flexibility for advanced chatbot workflows. Glide’s automatic model selection hides complexity from users.

These platforms represent a spectrum: from hyper-specialized no-code chatbots (OpenAI GPTs), to full-stack app coders (Bolt), to robust workflow orchestrators (Dust, Flowise), to general-purpose builders with AI extensions (Bubble, Glide), and finally, to end-user productivity agents (Cognosys). Each serves a unique audience with trade-offs between control, complexity, and capability.

\section{Trade-offs and Limitations}
While zero-code LLM platforms offer impressive capabilities, they also come with trade-offs. It is crucial to understand the limitations in terms of customizability, performance, scalability, and maintenance when relying on these higher-level tools. We discuss several key issues below.

\subsection{Limited Customizability vs. Flexibility}
By design, no-code platforms restrict the range of features and integrations to keep the experience simple. This can become a constraint when building applications that require specific logic, external integrations, or advanced reasoning capabilities. For instance, OpenAI’s GPT builder may not support custom tools or long-term memory, and agent platforms may prevent model fine-tuning or algorithmic changes. Some platforms address this through low-code options (e.g., Dust’s custom code modules or Flowise’s node customization), but these still have ceilings in complexity. While ideal for rapid prototyping and common use-cases, complex projects may demand code-centric solutions with full control over logic and infrastructure.

\subsection{Scalability and Performance}
No-code platforms introduce performance overheads and are not always optimized for scale~\cite{yellowNoCode2024}. Workflows involving multiple LLM calls may be slow or cost-inefficient at production scale. For instance, chaining 10 LLM calls in a loop for each user can become prohibitive in terms of latency and API costs. Platforms like Dust allow deployment via APIs (integrated into scalable web backends), while others like Bubble and Glide rely on their own cloud infrastructure, likely with built-in autoscaling. However, developers must still factor in LLM API limits, context window constraints, and real-time performance needs. High-volume or low-latency applications often require migration to optimized, code-based implementations.

\subsection{Vendor Lock-in and Data Ownership}
Reliance on proprietary no-code platforms risks vendor lock-in. Logic and data embedded in platforms like OpenAI GPTs or Cognosys cannot easily be exported or migrated. In contrast, Flowise (being open-source) mitigates this risk. Data residency and privacy are also concerns, especially when user queries or documents are processed by external LLM APIs. For sensitive use cases, organizations may need enterprise-grade features like private deployments or strict data governance policies. Trust in the provider’s data handling and continuity is essential.

\subsection{Quality and Reliability of AI Outputs}
LLMs can generate inaccurate or unpredictable outputs. When embedded in no-code applications, there’s limited support for validation or correction~\cite{zhang2023johnnycantprompt}. A Glide app extracting dates from unstructured text might fail subtly, with no built-in mechanism to verify the AI’s answer. Unlike traditional software, AI-driven logic lacks strong test frameworks. For critical applications, human oversight or fallback mechanisms are essential. Platforms rarely offer automated evaluation or fine-tuning on proprietary data, though some are exploring features for greater output reliability. As platforms abstract complexity, validating AI outputs becomes even more critical, highlighting the need for built-in evaluation pipelines—a limitation often overlooked in zero-code environments.

\subsection{Need for Prompt Engineering Skills}
Though labeled "no-code," many platforms still require users to craft effective prompts. The ability to design precise instructions (e.g., "act as a tax advisor and extract expense totals from receipts") influences output quality significantly. Non-technical users may face a learning curve, and vague prompts often yield poor results. Templates and examples help (e.g., OpenAI GPTs provide starting prompts; Glide uses guided fields), but prompt engineering remains a critical skill.

\subsection{Comparatively Shallow Learning}
Teams using no-code platforms may build working applications without deeply understanding LLMs, APIs, or software architecture. While this accelerates development, it may hinder long-term capacity to scale or adapt. If a no-code tool fails to meet future needs, organization may lack internal expertise to transition to custom solutions. For business-critical workflows, having at least one technical expert involved remains wise.

\subsection{Summary Perspective}
Zero-code LLM platforms boost productivity and accessibility, enabling fast iteration and user-friendly experiences. However, they trade off fine-grained control, flexibility, and robustness. These tools are ideal for prototyping, internal tools, or domains tolerant of occasional errors. As noted in industry reviews, they “lower the entry barrier, but problem-solving, architecture, and scaling still require expertise.” Many teams may benefit from starting with no-code and transitioning to custom implementations as requirements mature.

\section{Comparison with Traditional and Low-Code Development}
\begin{figure}[h] 
\centering 
\includegraphics[width=0.3\textwidth]{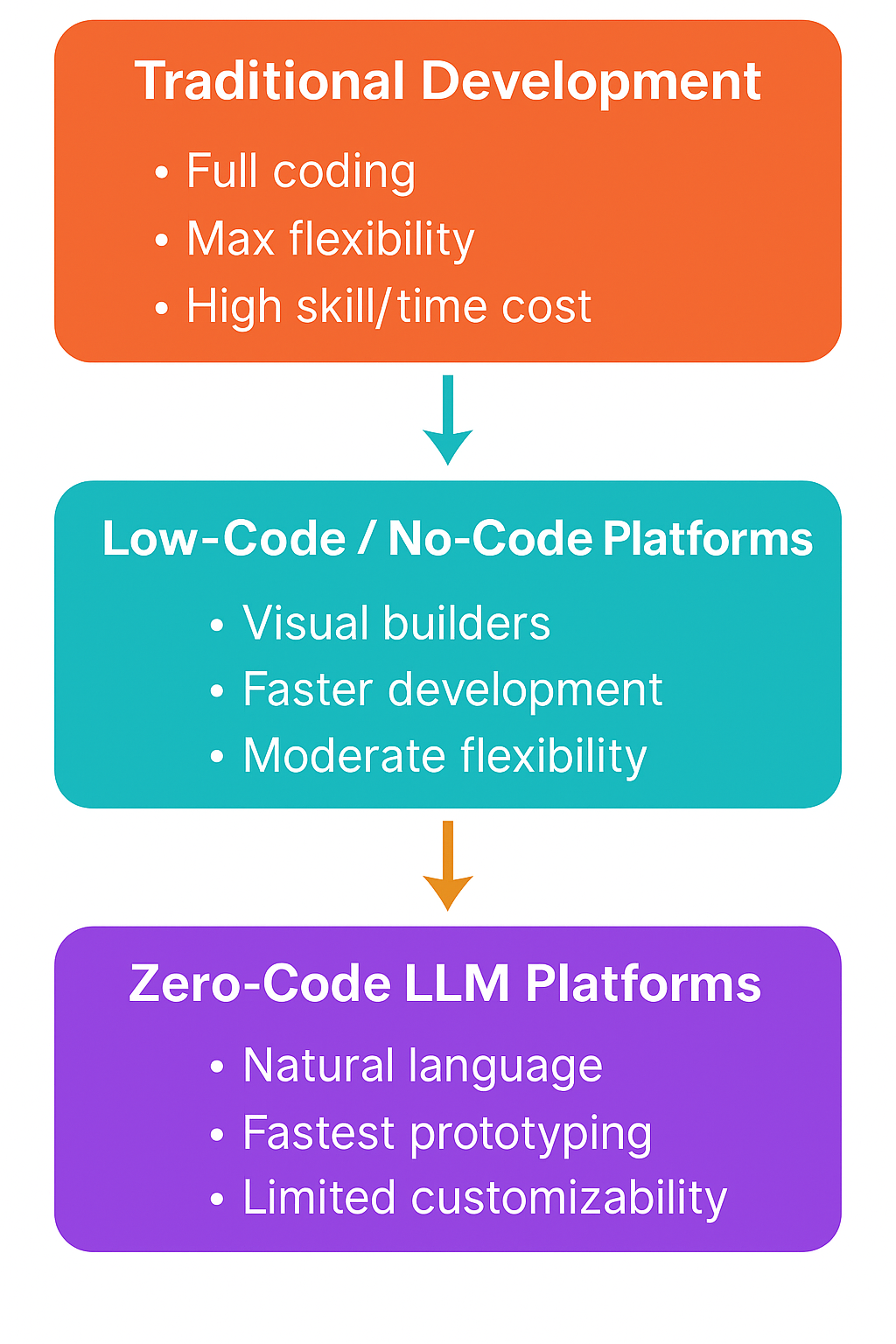} 
\caption{Development Paradigms: Traditional vs. Low-Code vs. Zero-Code LLMs.} \label{fig:dev-paradigms} 
\end{figure}

It is instructive to compare the capabilities of zero-code LLM platforms with both traditional software development and earlier low-code/no-code approaches~\cite{lowCode2020, noCodeSurvey2021, nocodecomparison2022} as shown in Fig \ref{fig:dev-paradigms}.

\subsection{Versus Traditional Coding}
Traditional development provides maximum flexibility~\cite{nocodecomparison2022}. Any desired functionality is achievable through explicit programming of the frontend, backend, and integrations. However, it is time-intensive and requires technical skill. Zero-code LLM platforms significantly reduce time to prototype. For instance, creating a Slack bot using an LLM and vector search may take days for a developer, but a non-technical user could build a similar solution in Dust or Flowise in hours~\cite{dustDocs2023, flowiseDocs}. The trade-off is control: developers can fine-tune embedding logic, add edge case handling, and optimize performance—capabilities not always available in no-code platforms.

No-code configurations may lack versioning, testability, or debugging tools found in traditional development~\cite{nocodetesting2021}. However, even professional developers may use LLM tools to avoid boilerplate, treating them as accelerators for repetitive patterns~\cite{copilot2021}. For complex, critical software, zero-code will not replace traditional methods, but for internal tools or simpler use-cases, it might fully suffice.

\subsection{Versus Prior Low-Code/No-Code (Without LLMs)}
Low-code platforms like Mendix or OutSystems allowed rapid visual app creation~\cite{mendix2020}. However, they struggled with non-standard logic unless supplemented with hand-written code. LLMs now fill that gap—natural language prompts can substitute complex logic blocks~\cite{zhang2023johnnycantprompt}. This elevates zero-code platforms to support scenarios that were previously unachievable without programming.

Whereas traditional no-code tools followed deterministic workflows, LLM-enhanced platforms are probabilistic: similar inputs may yield slightly varied outputs~\cite{varshney2023probabilisticnocode}. This adds complexity in testing and validation. Yet, LLMs make no-code accessible to domain experts through natural interaction. The shift from "drawing flowcharts" to "describing intent in language" represents a paradigm change~\cite{li2024languageUI}. Prompting is easier for many users than learning a visual programming interface.

\subsection{Coexistence and Integration}
Rather than a binary replacement, LLMs are being incorporated into traditional development. GitHub Copilot aids programmers inside code editors~\cite{copilot2021}. Microsoft's Power Platform Copilot brings LLM assistance into a low-code setting~\cite{microsoft2023copilot}. Builders can start with natural language and refine output via traditional editors.

In collaborative workflows, a domain expert might use a no-code LLM tool to build a first version, with engineers refining it for robustness or scale. This division of labor allows faster iteration and aligns responsibilities with expertise~\cite{gptworkflow2024}.

\subsection{When to Use What}
\begin{itemize}
    \item \textbf{Traditional coding:} Best for performance-critical, large-scale, or highly customized applications.
    \item \textbf{Low-code:} Suited for structured apps with moderate custom logic.
    \item \textbf{Zero-code LLM:} Ideal for fast prototyping, internal tools, or automating domain-specific tasks by non-programmers.
\end{itemize}

A marketing team could independently build a social media assistant with a no-code LLM tool, saving engineering resources. Later, if it becomes mission-critical, developers might reimplement it for control and efficiency~\cite{nocodeevaluation2023}.

\subsection{Summary}
Zero-code LLM platforms complement rather than replace traditional development. They unlock rapid iteration, empower non-developers, and reduce prototyping costs. However, they trade off control, reliability, and scalability~\cite{zhang2023johnnycantprompt, nocodeevaluation2023}. The future likely lies in a hybrid model: fast prototyping via LLM-enhanced no-code tools, followed by productionization by engineers. As industry analysis suggests, these platforms "streamline development" but don’t eliminate the need for engineering expertise~\cite{yellowNoCode2024}.

\section{Future Directions}
The field of no-code LLM platforms is evolving rapidly. We highlight several trends and future directions that are likely to shape the next generation of tools in this space~\cite{nocodeevaluation2023, yellowNoCode2024, gptworkflow2024, teamagent2024, openai2023gpts, dustDocs2023, li2024languageUI, mistral2023models, monsterapi2023, wang2024debugging, fan2023toolllm}.

\subsection{Multimodal Inputs and Outputs}
Future no-code platforms will likely move beyond text~\cite{gpt4vision2023, dalle2022, whisper2022}. Some platforms already support image (OCR/description) and audio (speech-to-text) inputs. As multimodal LLMs become widespread, creators will build apps that process voice, images, or even video. For instance, a mobile app could let users speak a query and receive a video or AR output generated by connected AI models. Tools like Dust.tt already integrate visualization capabilities~\cite{dustDocs2023}, and we expect further adoption of DALLE, Stable Diffusion, and text-to-speech engines. Users may soon assemble voice+vision+text apps entirely through drag-and-drop interfaces.

\subsection{On-Device and Private LLMs}
Model compression and stronger edge devices make it possible to run LLMs locally~\cite{mistral2023models}. Future no-code platforms may allow on-device LLMs for privacy or offline usage. For example, a mobile app builder might ship with a local LLM for secure, fast inference. Enterprise tools may offer one-click deployment of open-source models to private clouds. We also anticipate no-code fine-tuning options for domain-specific models (e.g., via UIs from MonsterAPI or H2O.ai)~\cite{monsterapi2023, h2o2023}. This will require hiding the complexity of training and deployment while retaining control.

\subsection{Better Orchestration and Tooling}
LLM workflows today use basic chains or conditionals. Tomorrow’s no-code tools may include multi-agent orchestration: users could drop agents (research, write, critique) onto a canvas and define interactions~\cite{teamagent2024}. Debugging tools like trace visualizations or AI critic-review loops may become standard~\cite{wang2024debugging}. Expect safety features such as permissions toggles and automatic retry logic. This mirrors advancements in software engineering (e.g., versioning, monitoring, CI/CD) now ported to no-code AI stacks~\cite{fan2023toolllm}.

\subsection{Collaboration and Community Sharing}
Platforms are moving toward shared agent templates and galleries (e.g., OpenAI's GPT store~\cite{openai2023gpts}, Dust.tt’s template gallery~\cite{dustDocs2023}). A user could drag a prebuilt sales assistant bot into their app and tweak it. These templates may include prompt strategies, model settings, and integration configs. Community forums and marketplaces will reduce the barrier for entry and accelerate onboarding for non-experts~\cite{yellowNoCode2024}.

\subsection{Convergence with Traditional IDEs}
Just as visual editors appeared in code IDEs and code windows in no-code tools, we expect convergence~\cite{gptworkflow2024}. For instance, LangChain plugins for VS Code might visualize prompt graphs. Conversely, no-code tools may expose source views editable by developers. This duality allows mixed teams to collaborate using different interfaces on the same underlying project definition (e.g., in JSON synced with Git).

\subsection{AI Improvements Benefiting No-Code}
As LLMs improve—less hallucination, larger contexts, better reasoning—no-code tools inherit these benefits~\cite{li2024languageUI}. For example, longer context windows reduce the need for chunking knowledge. Model ensembles (intent classifier + generator + checker) can be abstracted from users while providing better quality~\cite{fan2023toolllm}. Prompt understanding will improve, making verbal app descriptions a viable way to specify full logic flows.

Zero-code LLM platforms are moving toward higher capability and accessibility. Multimodal input/output, local model support, smarter orchestration, and tighter integration with coding workflows will redefine what's possible. As these platforms evolve, the line between developer and non-developer tools will blur, making "AI-assisted development for all" a tangible reality.

\section{Conclusion}
Zero-code LLM platforms represent a major step toward democratizing software creation. They empower wide range of users—from professionals to hobbyists—to build intelligent applications using natural language and visual interfaces. In this survey, we analyzed leading platforms (OpenAI GPTs, Bolt.new, Dust.tt, Flowise, Cognosys, Bubble, Glide), provided a taxonomy for classification, and discussed capabilities like agent autonomy, memory, integration, and extensibility. We found that while these platforms differ in approach, they all aim to make LLM-powered development accessible without coding. Their impact is visible in faster prototyping cycles, broader participation in app creation, and new workflows for collaboration of domain experts and engineers.

However, challenges remain. These tools trade off control, scalability, and validation capabilities for accessibility and speed. Concerns like vendor lock-in, reliability of AI outputs, and lack of deeper software understanding persist. Yet, their future is promising. As AI improves, we anticipate platforms integrating multimodal input/output, on-device deployment, agent orchestration, and hybrid code-no-code collaboration features. In time, zero-code LLM platforms may become the default for building AI-powered solutions—enabling conversational, intelligent development that lowers barriers while maintaining quality and flexibility. The dream of "software by anyone" is closer than ever.

\bibliographystyle{IEEEtran}
\bibliography{citations}

\end{document}